\documentclass[11pt,a4paper]{article}
\usepackage{amsmath}
\usepackage{amscd}
\usepackage{amsthm}
\usepackage{amssymb} \usepackage{latexsym}
\usepackage{graphicx}
\def\myscale{0.9}
\newcommand{\bel}[1]{\begin{equation}\label{#1}}

\newcommand{\be}{\begin{equation}}
\newcommand{\qe}{\end{equation}}
\newcommand{\ba}{\begin{eqnarray}}
\newcommand{\ea}{\end{eqnarray}}
\newcommand{\rf}[1]{(\ref{#1})}
\newcommand{\bi}{\bibitem}
\date{November 5, 2003}

\begin{document}
\title{
On the emergence of complex systems on the basis of the coordination
of complex behaviors of their elements
 \\
\vspace{2.5cm}
}
\author{
  Fatihcan M. Atay\footnote{Max Planck Institute for
    Mathematics in the 
  Sciences, Inselstr.~22-26, 04103 Leipzig, Germany, \{fatay, jost\}@mis.mpg.de},
   J\"urgen Jost\footnotemark[\value{footnote}]
  			  \footnote{Santa Fe 
  Institute, 1399 Hyde Park Road, Santa Fe, NM 87501, USA, jost@santafe.edu}}
\maketitle
\abstract{{\sl We argue that the coordination of the activities of
  individual complex agents enables a system to develop and sustain
  complexity at a higher level. We exemplify relevant mechanisms
  through 
  computer simulations of a toy system, a coupled map lattice with
  transmission delays. The coordination here is achieved through the
  synchronization of the chaotic operations of the individual
  elements, and on the basis of this, regular behavior at a longer
  temporal scale emerges that is inaccessible to the uncoupled individual
  dynamics.}}
\bigskip

The purpose of this article is to challenge the view, often expressed
and perhaps prevalent in most discussions, that the essence of complex
systems lies in the emergence of complex structures from the
non-linear 
interaction of many simple elements that obey simple rules. Typically,
these rules consist only of
 0-1 alternatives selected in response to the input received, as in
 many prototypes like cellular automata, Boolean networks, spin
 systems, etc. We do not intend to deny that quite intricate patterns
 and structures can occur in such systems. However, these are toy
 systems, and the systems occurring in reality rather consist of
 elements that individually are quite complex
 themselves.\footnote{Throughout this essay, we employ the term
   ``complex'' only in some vague and metaphorical manner, without any
   attempt at quantifying it. This should not obscure the general
   thesis presented here.} This brings
 in a new aspect that seems essential and indispensable to the
 emergence and functioning of complex systems, namely the coordination
 of individual agents or elements that themselves are complex at their
 own scale of operation. This coordination dramatically reduces the
 degrees of freedom of those participating agents. Understanding the
 mechanisms responsible for 
 achieving and maintaining this coordination seems the key to
 understanding, for example, the major transitions in evolution
 \cite{MSS}. Even the 
 constituents of molecules, the atoms,  are rather
 complicated conglomerations of subatomic particles, perhaps
 ultimately excitation patterns of superstrings. Genes, the elementary
 biochemical coding units, are complicated macromolecular strings, as
 are the metabolic units, the proteins. Neurons, the basic elements of
 cognitive networks, themselves are cells. While their activity
 follows an apparently simple pattern of firing vs. resting, this
 depends on a slower learning dynamics tuning the strengths of the
 synaptic connections between them according to the history of
 temporal correlations between pre- and postsynaptic activities. At an
 even higher level of aggregation, an economic system consists of the
 interaction of humans, obviously highly complex agents. Nevertheless,
 standard economic theory is rather successful even though it
 assumes that these agents follow quite simple rules as laid down in
 utility functions and optimization patterns.
 
In any of these examples, it is by no means evident that the
interactions of the elements or agents leads to a coherent structure
at a higher level. If you bring a heterogenous group of people
together, they will not automatically build a smoothly functioning
economic system. It is rather that a functioning economic system has
some subtle means to suppress the individual and disruptive behavior
of its members and coerce them to operate in a manner that to a
sufficiently large degree is
predictable for the others. The rules and institutions that guarantee
the functioning of the economic system are either directly imposed 
like the legal framework of contracts and the monetary system and then
adapted by the economic system according to its internal exigencies,
or acquired by the participants through processes of socialization,
education, and experience. It is not our purpose here to enter the
ongoing debate to what extent the rationality assumptions underlying
standard economic theory are justified when contrasted with empirical
investigations of the behavior of individual economic agents. We
rather wish to make the point that economic agents behave rationally
to whatever degree they do so because and to the extent to which they
are participants in an economic system. Turning to another one of our
examples, the behavior of neurons in vivo is different from the one in
vitro, the former one exhibiting more regularities that are not
intrinsic to the operation of the individual neuron itself, but rather
imposed by the neural system in which the neuron is participating. In
other words, in this and 
other complex 
systems, it is an important feature that the potential complexity of
the behavior of the individual agents gets dramatically simplified
through the global interactions within the system. The individual
degrees of freedom are drastically reduced, or, in a more formal
terminology, the factual state space of the system is much smaller
than the product of the state spaces of the individual elements. This
is one key aspect. The other one is that on this basis, that is
utilizing the coordination between the activities of its members, the
system then becomes able to develop and express a coherent structure
at a higher level, that is, an emergent behavior that transcends what
each element is individually capable of. Our thesis then is that the
essence of a theory of complex systems should rest in analyzing and
understanding the interplay of those two aspects. The reduction of the
individual possibilities opens new possibilities at a higher level.

For a deeper conceptual analysis, one should then consider the
elements or agents not as part of the system, but rather as
constituting an inner or interior environment for the system, as in
\cite{Lu}, so as to focus on the principally irreducible context of
the system level. Here, however, rather than pursueing these
conceptual aspects (see \cite{J1,J2} in that direction), we wish to
elucidate this through a formal model system. As 
argued, such a system should not consist of simple agents, but rather
ones that already by themselves possess a certain degree of
complexity. We choose a discrete time chaotic dynamical system, namely
the iteration of 
the logistic map
\bel{1}
f(x)=\rho x(1-x)
\qe
with a parameter $1 \le \rho \le 4$. The iteration proceeds via
\bel{2}
x(n+1)=f(x(n))
\qe
for some starting value $x(0)$ ($n \in \mathbb{N}$). 
$f=f_\rho$ maps the unit interval
$[0,1]$ to itself. As we let $\rho$ increase towards 4,
periodic orbits appear through successive period doubling bifurcations
until the behavior eventually becomes fully chaotic. Since the
iteration of $f_\rho$ for $\rho$ 
sufficiently close to 4 amplifies small differences
of the starting values, the future of an iteration cannot be predicted
unless one makes the unrealistic assumption that the starting value
is known with infinite precision.  See \cite{ASY} or a similar
textbook for an introduction. 

Our system couples such individual chaotic dynamical systems. We
assume that we have some graph $\Gamma$. Vertices $x,y$ connected by an edge
are called neighbors, symbolically denoted by $x \sim y$. The number
of neighbors of $x$ is denoted by $n_x$. For a
parameter $\epsilon$, the coupling leads to the system
\bel{3}
x(n+1)= f(x(n))+ \frac{\epsilon}{n_x} \sum_{y \sim x} (f(y(n))-f(x(n))).
\qe
Thus, $x$ now adjusts its  state not only the basis of its own
present state, but also takes the state differences from its neighbors
into account. The coefficients on the right hand side are chosen in
such a manner that the total weight of all the contributions is 1,
that is, the same as in \rf{2}. 
This a coupled map lattice as introduced and studied for fully
connected graphs by Kaneko \cite{Ka1,Ka2}. In particular, he discovered the
phenomenon of synchronization of chaos, that is for certain values of
$\epsilon$ and certain graphs, the individual chaotic iterations
operate synchronously, that is
\bel{4}
x(n)=y(n) =: \Xi(n)
\qe
for all vertices $x,y$ of the graph and for sufficiently large $n$,
regardless of the different starting values for the iterations at the
individual nodes. In other cases, one may also observe intermittent
behavior, that is, synchrony goes on and off. Mathematically, the
stability of the synchronized solution can be studied through
perturbations by eigenfunctions of the graph  Laplacian
\bel{5}
\Delta \phi(x):=  \frac{1}{n_x} \sum_{y \sim x} (\phi(y)-\phi(x)),
\qe
and this is the reason why we prefer to write the system as in \rf{3}
instead of in the apparently simpler form
\bel{6}
x(n+1)= (1-\epsilon)f(x(n))+ \frac{\epsilon}{n_x} \sum_{y \sim x} f(y(n)).
\qe
See for example \cite{JJ} for an analysis. Whether synchronization
occurs depends essentially on the spectral gap of the graph $\Gamma$, that is
on the value of the first non-trivial eigenvalue of
$\Delta=\Delta_\Gamma$ (which in turn reflects the topology, that is,
the connection structure of the underlying graph $\Gamma$), and, of
course, on the coupling parameter $\epsilon$. 

Synchronization is perhaps the most basic mechanism for the
coordination of the behavior of individual elements or agents whose
intrinsic dynamics are coupled. See \cite{PRK} for a general
introduction. Synchronization dramatically reduces the degrees of
freedom for the dynamics of the coupled system when compared to the
uncoupled dynamics of the individual agents, inasmuch as the
synchronized dynamics is fully characterized by the dynamics of a
single element. 

The situation described so far is one where the synchronized
collective dynamics coincides with the individual dynamics of an
element in the uncoupled state, and so can be predicted by the
latter. In particular, this only corresponds to the first one of the
two key aspects for the emergence of complex behavior that we
identified above. In order to obtain a new type of collective
dynamics, we need to introduce an additional feature. Following
\cite{AJW}, the feature we choose is a temporal delay in the
transmission of the activities between vertices. In formal terms, we
consider the coupled system
\bel{7}
x(n+1)= f(x(n))+ \frac{\epsilon}{n_x} \sum_{y \sim x}
(f(y(n-d_{yx}))-f(x(n))) 
\qe
where $d_{yx} \in \mathbb{N}$ is the delay\footnote{In the sequel, we
  shall only consider constant transmission delays, $d_{yx}\equiv d$.}
from vertex $y$ to 
$x$. This leads to several new dynamical features that we shall
describe and explore in more detail and utilize to support the
paradigm developed above. On one hand, we can generate synchronized --
chaotic or regular -- 
behavior that is different from the chaotic dynamics \rf{1}, \rf{2} of
an uncoupled element. On the other hand, we can also generate
regularities on a longer time scale that transcend the capabilities of
isolated elements. On a technical level, we can even sustain period 3
oscillations stably over some parameter range, in contrast to the fact
that for an isolated dynamical iteration, this is the penultimate
state before chaos sets in \cite{Sha,LY}.
 
The first observation relevant here is that the uncoupled dynamics
\rf{2} is no longer a solution of \rf{7}, in contrast to the system
\rf{3} without delays. In the
simplest possible case of global synchronization, the collective
behavior can be obtained 
through a temporal averaging from individual dynamics, but in more
interesting cases, it is fully irreducible.  Thus, even the dynamics
of a collective 
synchronized behavior cannot be reduced to the individual dynamics
anymore, but rather reflects a new collective system dynamics.

We now describe some of the simulation results in more detail; the
mathematical treatment will be given elsewhere. We consider a
scale-free graph with 10,000 nodes.The results for random graphs are
qualitatively similar, whereas for regular graphs with
nearest-neighbor coupling, synchronization is typically not observed,
see \cite{JJ}  for the case without delays where this behavior finds an
explanation from the properties of the spectrum of the graph
Laplacian.

\begin{figure}[tb]
\begin{center}
\includegraphics[scale=\myscale]{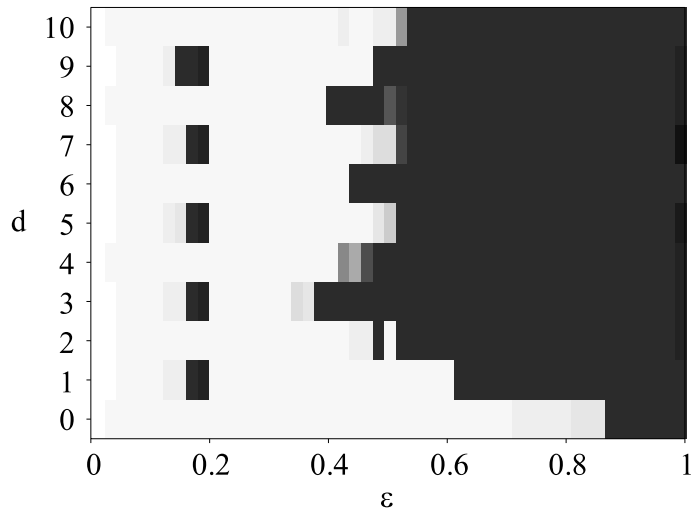}
\caption{ \label{fig:scale-free}
Synchronization of a scale-free network. 
The gray scale shows the degree of synchronization,
with black corresponding to full synchronization.}
\end{center} 
\end{figure}

We start with the case 
that is maximally chaotic in the uncoupled case, namely $\rho=4$. 
Figure~\ref{fig:scale-free} indicates through gray values for which values of the
coupling parameter 
$\epsilon$ and constant delay $d$ the dynamic synchronize. In
particular, the system synchronizes more readily, that is, starting at
lower values of $\epsilon$ in the presence of delays than
without. Also, there is a critical region roughly between $\epsilon
=.1$ and $.2$ where synchronization occurs for odd, but not for even
delays.

\begin{figure}[tb]
\begin{center}
\includegraphics[scale=1.3]{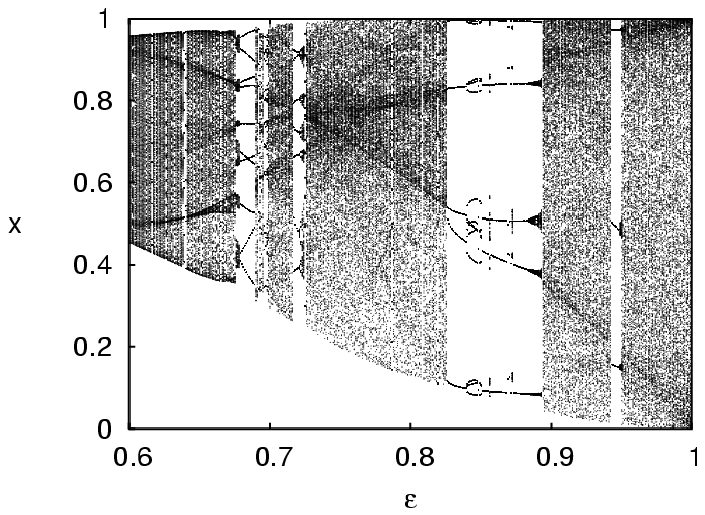}
\caption{\label{fig:bifur_eps}
The dependence of the synchronized solution on the coupling parameter $\epsilon$.
}
\end{center} 
\end{figure}

We now explore some of these effects in more detail. We
consider constant delay $d=1$ for the transmission between vertices,
and we display the behavior of the coupled and delayed dynamics as
depending on the coupling parameter $\epsilon$. 
Figure~\ref{fig:bifur_eps} is a
bifurcation diagram, 
exhibited in the range $\epsilon > 0.6$ for which synchronization is observed.
 As $\epsilon$ varies, we see chaos intermittent with
 periodic behavior; period 5 is quite stable in the range of
 $\epsilon$ between .8 and .9, while we see period 3 around
 $\epsilon=.94$. Periodic solutions become rarer for larger $d$.

\begin{figure}
\begin{center}
\includegraphics[scale=\myscale]{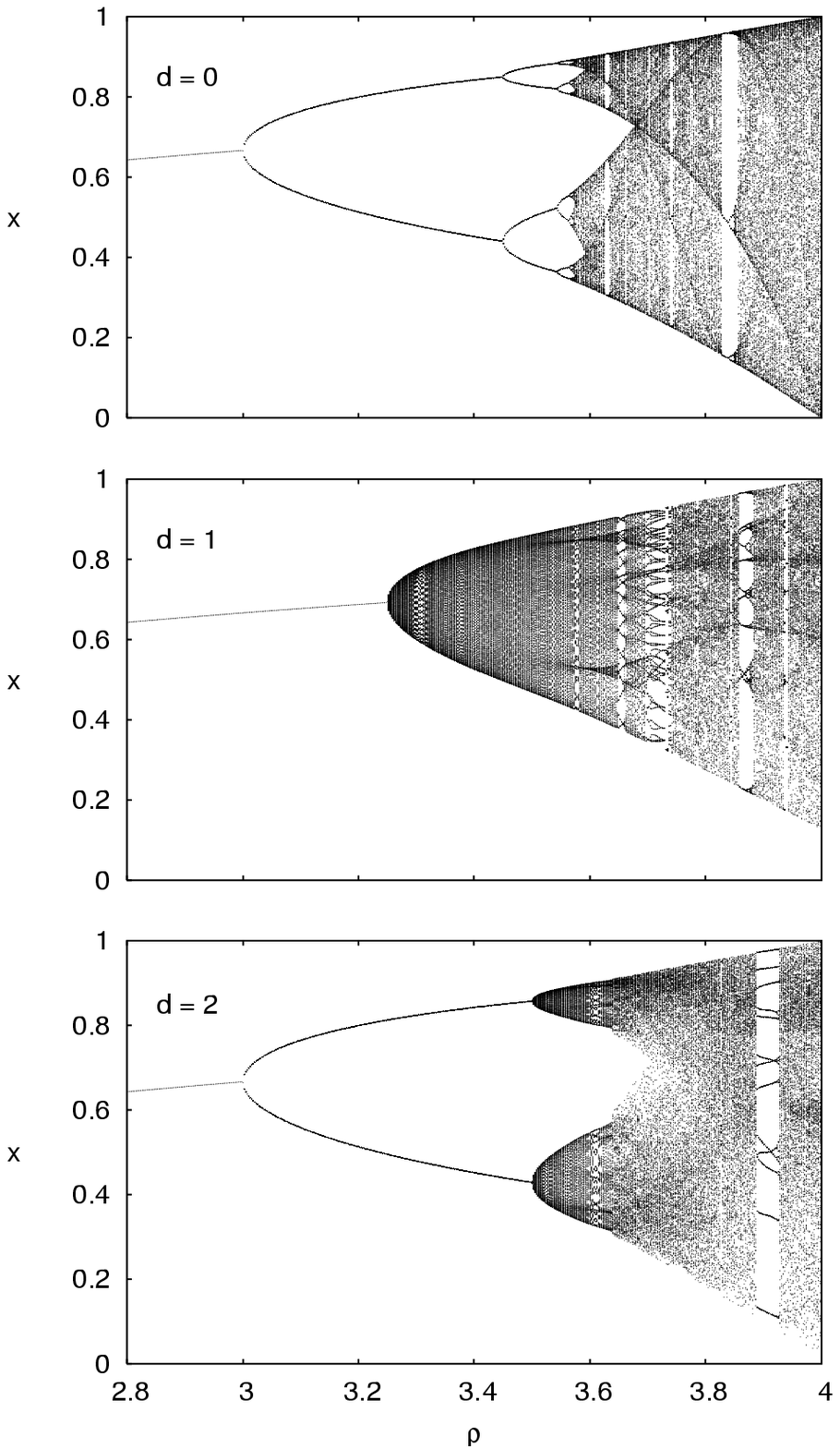}
\caption{\label{fig:bifur_rho}
The dependence of the synchronized solution on the parameter $\rho$.
}
\end{center} 
\end{figure}

We next fix $\epsilon$ at .8 which is within the synchronization
region, take $d=1$ (the results for other odd $d$ are similar), and
let $\rho$ increase from 2.8 to 4. We find a constant solution up to
$\rho \approx 3.2$ and then a direct transition to high period
solution without intermediate successive period doublings (Figure~\ref{fig:bifur_rho}). 
Thus, the route to chaos is
different here from the standard period doubling paradigm. Another
important difference is that we now get two positive Lyapunov
exponents instead of one in the undelayed case, that is, we see
a higher level of complexity at the system level than could be
sustained by the individual dynamics.
\begin{figure}
\begin{center}
\includegraphics[scale=\myscale]{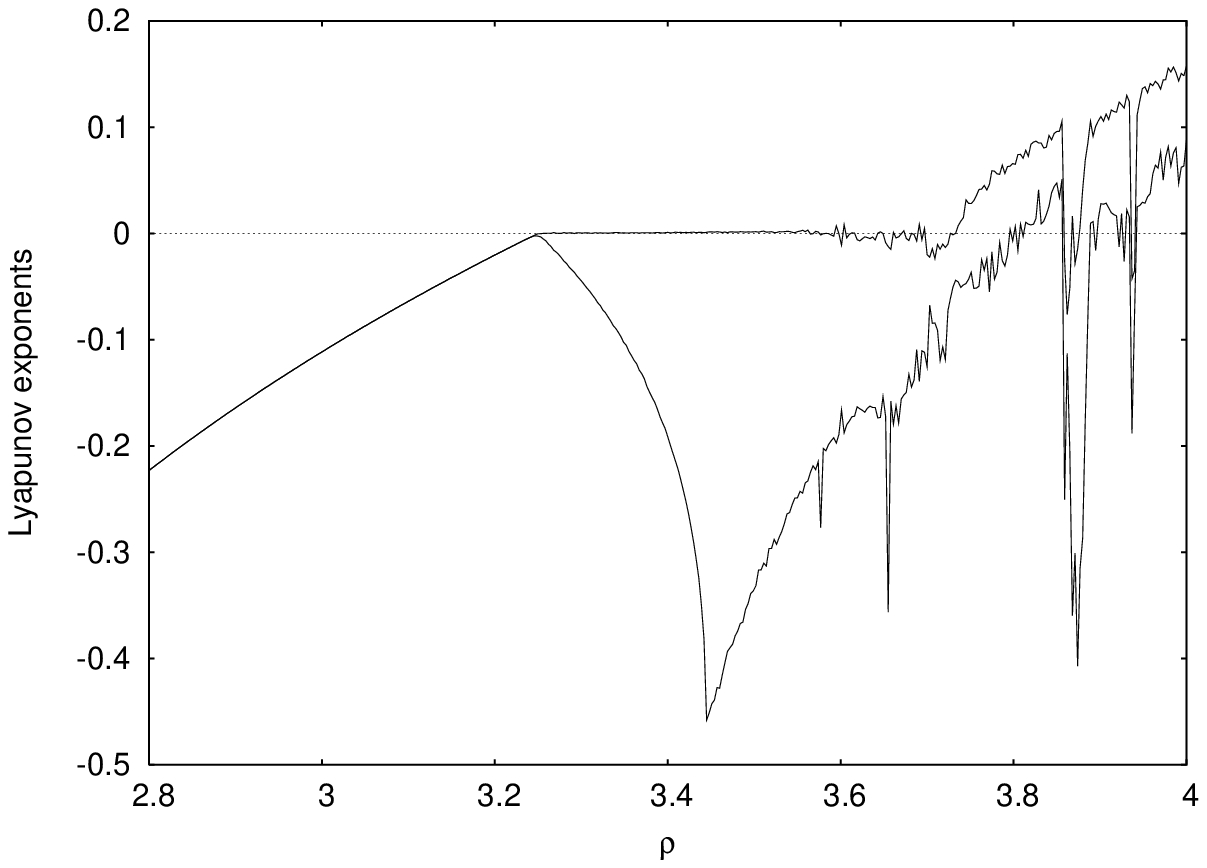}
\caption{\label{fig:lyap}
The Lyapunov exponents of the synchronized solution calculated for $d=1$ and
$\epsilon=0.8$.
}
\end{center} 
\end{figure} 
For $d=2$,
we see one period doubling at $\rho \approx 3$ which is also the value
near which period doubling occurs for the standard uncoupled logistic
map. In contrast to the latter, however, we do not observe further
period doublings, but rather a Neimark-Sacker bifurcation at $\rho
\approx 3.45$. That means that we get a pair of complex conjugate
eigenvalues crossing the unit circle, but in contrast to the
standard Hopf bifurcation for continuous time dynamics, here high
period solutions bifurcate from a 
fixed point. For $d=4$, the behavior is similar, but we see two
consecutive period doublings before the Neimark-Sacker bifurcation.

\begin{figure}
\begin{center}
\includegraphics[scale=\myscale]{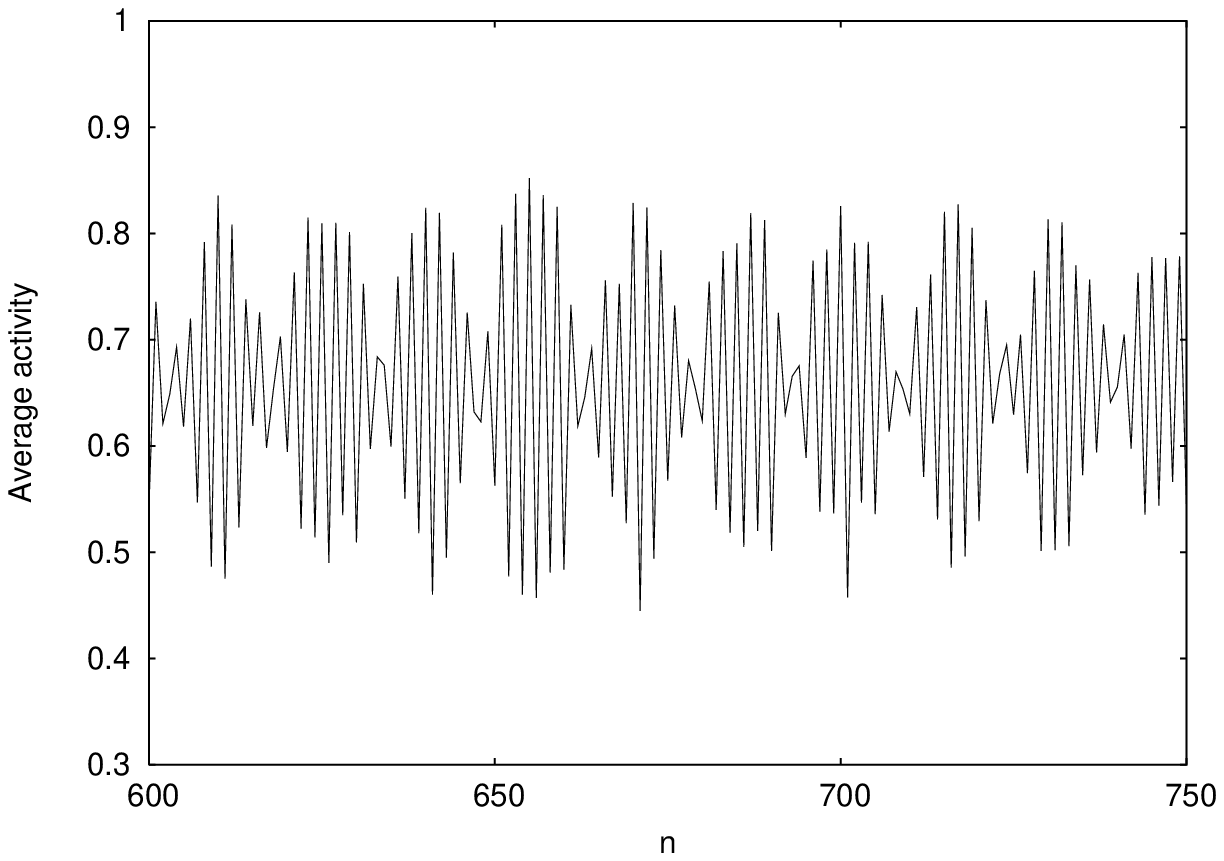}
\caption{\label{fig:envelope}
The average activity of the network for a non-synchronized solution, 
obtained for $d=8$, $\rho=4$, and $\epsilon=.135$. 
}
\end{center} 
\end{figure}

While all this is technically, but perhaps not so much qualitatively
differently from the chaotic behavior of the uncoupled logistic
dynamics, in the region between the $\epsilon$-values .1 and .2,
already mentioned above as yielding different behavior depending on
the parity of $d$, we find a qualitatively different behavior for
larger even values of $d$, namely, a non-synchronized region with an
enveloping curve of the dynamics that shows long time periodic
behavior (Figure~\ref{fig:envelope}). 
This behavior is only seen in the collective dynamics, but
not the individual one, and it occurs on a longer time scale than
accessible to the latter. 

Obviously, one can find emergent collective dynamics in other coupled
systems, liking networks of spiking neurons, see for example
\cite{GK}. In most such cases, however, this type of behavior is
caused by underlying stochastic effects. The framework exhibited here
offers some possibilities for a direct analytic approach to
understanding such phenomena (see \cite{AJW} in this direction). When
compared with the conceptual setting described in the beginning, one
deficit of the present model is perhaps  that the coupling parameter
$\epsilon$ has to be set by hand, instead of self-emerging from the
intrinsic dynamics of the system. Also, the setting here is completely
unrealistic in the sense that no individual variations are allowed,
that is, all elements operate with the same parameter
values,\footnote{It is not principally difficult, however, to extend
the model and the simulations to cases where individual variations
of the parameters like $\rho$ are allowed.} with the
only exception that the underlying graph structure is not uniform or
regular, but random. Nevertheless, looking at our
simulation results, hopefully some understanding can be gained for our
thesis that higher level complex behavior depends on the coordination
of the activities of the participating agents which are complex
themselves. As long as these operate in an uncoordinated manner, no
higher scale is available for the encompassing system.

In any case, our formal model on which the simulations are based is
obviously woefully inadequate to reflect the richness of the examples
of higher level complex systems quoted. It shares this deficit with
the formal models mentioned in the beginning of this essay. We think,
however, that the present model captures one important aspect that is
not represented in those ones. We also hope that a deeper formal
analysis of that aspect will yield further insights into the
mechanisms leading to the emergence of higher level complex systems.

\bigskip

{\bf Acknowledgement:} The computer code for the simulations displayed
here was written by Andreas Wende.

\end{document}